\documentclass[aps,twocolumn]{revtex4}
\def\[{\left\lbrack}
\def\]{\right\rbrack}

\def\({\left(}
\def\){\right)}
\newcommand{\be}{\begin{equation}}
\newcommand{\ee}{\end{equation}}
\newcommand{\ea}{\end{eqnarray}}
\newcommand{\ba}{\begin{eqnarray}}

\begin{document}

\title{Clebsch parameterization from the symplectic point of view}

\author{C. Neves and W. Oliveira}
\thanks{\noindent e-mail:cneves, wilson@fisica.ufjf.br}
\affiliation{Departamento de F\'{\i}sica, ICE, Universidade Federal de Juiz de Fora,\\
36036-330, Juiz de Fora, MG, Brasil}

\begin{abstract}
\noindent
This work propose an alternative and systematic way to obtain a canonical Lagrangian formulation for rotational systems. This will be done in the symplectic framework and with the introduction of extra variables which enlarge the phase space. In fact, this formalism provides a remarkable and new result to compute the canonical Lagrangian formulation for rotational systems, {\it i.e.}, the obstruction to the construction of a canonical formalism can be solved in an arbitrary way and, consequently, a set of dynamically equivalent Lagrangian descriptions can be computed. 
\end{abstract}

\maketitle

\setlength{\baselineskip} {20 pt}

\section{Introduction}

Recently, fluid mechanics\cite{LL} has attracted a huge attention to theoretical physicists\cite{HB,JB} due to the investigate of how some instances of the classical theory are related to D-branes and how this relation explains some integrability properties of several models. In Ref.\cite{HB}, the authors demonstrated that the (d + 1)-dimensional relativistic theory of D-branes are integrable systems by reducing the problem to a d-dimensional irrotational fluid mechanics. Afterwards, Bazeia and Jackiw\cite{JB}, found the solutions of this Galileo invariant system that are in connection with the solutions of the relativistic D-brane system. In particular, these works clarify the presence of a hidden dynamical Poincar\'e symmetry on the d-dimensional fluid mechanics. However, this is only valid only when the rotational fluid model has a specific potential, $(V\propto 1/\rho$ with $\rho$ as being the mass density).

In this paper, we propose an alternative answer to the question on the Lagrangian for velocities fields that are not irrotational. It is known that for rotational fluid mechanics, whose vorticity nonvanishes, the symplectic two-form does not exist; it happens because its respective \lq\lq inverse\rq\rq is singular and, consequently, has a zero-mode given by the gradient of a quantity called \lq\lq Casimir invariants\rq\rq. Since these quantities  Poisson-commute with all dynamical variables and Hamiltonian and are also constant of motion, there is no symmetry related. As demonstrated by Jackiw in Ref.\cite{jackiw}, the algebra that admits Casimir invariants also creates an obstruction to the construction of a canonical formalism for fluid mechanics. Indeed, this obstruct the determination of the fluid dynamical Lagrangian. This obstruction was neutralized by C.C. Lin\cite{CCL} using the Clebsch parameterization, which was discussed by Jackiw\cite{jackiw} recently.

In order to achieve our goal and present a self contained paper, this work is organized as follows. In the next section, we present the general formalism that implement the Clebsch parameterization from the symplectic point of view. In order to illustrate and familiarize the reader with the problem proposed and its solutions given by the Clebsch-symplectic parameterization process, we present in the Section III the rotational fluid system as well as the obstruction problem. In Section IV, we apply the Clebsch-symplectic parameterization formalism to the rotational fluid system in order to solve the obstruction problem. In consequence, we obtain the Lagrangian for the rotational system. Further, we also discuss the arbitrarity present on the determination of the rotational fluid Lagrangian. In fact, we can exhibit a set of different, but dynamically equivalent Lagrangian descriptions for the rotational fluid mechanics. In the last section, we express our find outs and conclusions.

\section{General formalism}

In order to systematize the Clebsch-symplectic parameterization formalism, let us consider a mechanical system governed by a Lagrangian $({\cal L}\equiv{\cal L}(q_i,\dot q_i, t)$ with $i,j=1,2\dots, N$, where dot means temporal derivative) and whose vorticity nonvanishes, {\it i.e.}, ${\vec\omega}=\bigtriangledown\times\dot{\vec q}\not=0$.

First all, the Lagrangian must be rewritten into its first-order form, namely, ${\cal L}^{(0)} = a_i\dot\xi^i - H(\xi)$, where $H(\xi)$ is the Hamiltonian and $\xi^i$ are the symplectic variables. The Euler-Lagrange equation of motion are $f_{ij} \dot\xi^j = \frac{\partial H(\xi)}{\partial \xi^i}$, where $f_{ij} = \frac{\partial a_j(\xi)}{\partial \xi^i} - \frac{\partial a_i(\xi)}{\partial \xi^j}$.
Note that the equation of motion above for $\xi$ is well defined if the matrix $f_{ij}$ possesses the inverse $f^{ij}$, {\it i.e.}, $\dot\xi^j = f^{ji} \frac{\partial H(\xi)}{\partial \xi^i}$. From the Hamilton approach, the equation of motion can be expressed bracketing the variables with the Hamiltonian, $\dot\xi^j = \{\xi^j,H(\xi)\} =\{\xi^j,\xi^i\}\frac{\partial H(\xi)}{\partial \xi^i}$. Now, we are led to postulate the fundamental brackets as $\{\xi^j,\xi^i\} = f^{ji}$. When $f_{ji}$ has no inverse, {\it i.e.}, $f_{ji}$ is singular and, subsequently, constraints arise or there are gauge symmetries present within the system, which were well investigated and solved by Barcelos and Wotzasek\cite{BW}. However, an other obstacle can appear. This obstacle arises due to the existence of a quantity $C(\xi)$ whose Poisson bracket with all symplectic variables $\xi^i$ vanishes, $0 = \{\xi^j, C(\xi)\} = f^{ji}\frac{\partial C(\xi)}{\partial \xi^i}$. This means that $\frac{\partial C(\xi)}{\partial \xi^i}$ is a zero-mode of $f^{ji}$ and its inverse , the symplectic two-form $f_{ji}$, does not exist. Since these quantities $C(\xi)$ Poisson-commute with all variables and Hamiltonian then they are also constants of motion, and are called \lq\lq Casimir invariants\rq\rq. It is important to notice that these constants do not reflect any symmetries and that they do not generate infinitesimal transformations. In view of this, the existence of Casimir invariants create an obstruction to the construction of a canonical formalism for fluid mechanics and the Lagrangian description for this system is also frustrated. To overcome this problem and develop to a more advanced stage, this obstruction must be neutralized. It was firstly done by C.C. Lin\cite{CCL} through Clebsch parameterization, which bases on vector properties. Here, we have solved the obstruction generated by Casimir invariants using the symplectic formalism. In consequence, we determine the dynamical Lagrangian for the rotational system. It is important to notice that a set of dynamical equivalent Lagrangian descriptions for the rotational system can be determined.

The Clebsch-symplectic parameterization formalism has its bases on the introduction of extra variables embraced by arbitrary functions, $\Psi\equiv\Psi(q_i,a_i,\alpha,\beta)$ and $G\equiv G(q_i,a_i,\alpha,\beta)$, where the extra variables $\alpha$ and $\beta$ respect the following algebra $\{\alpha, \beta\} = A$, with $A$ being an arbitrary parameter defined conveniently, and $G$ is a function expanded in terms of $\eta=(\alpha, \beta)$, $G=\sum_{n=1}^\infty {\cal G}^{(n)}(\bar\xi), \hspace{.5cm}{\text with}\hspace{0.3cm} {\cal G}^{(n)}(\bar\xi)\propto \eta^n$, and $\bar\xi^k=(q_i,a_i,\alpha,\beta)$ with $k=1,2,\dots, n+2$. The $G$ function satisfies the following boundary condition $G(\xi^i, \eta=0)=0$.

In order to begin with this alternative parameterization process, these arbitrary functions are introduced into the first-order Lagrangian, which becomes ${\bar{\cal L}}^{(0)} = \bar a_i\dot{\bar\xi^i} + \Psi\dot\beta - {\bar H}(\bar\xi)$,where ${\bar{\cal H}}(\bar\xi) = H(\xi) + G(\bar\xi)$.
The symplectic two-form, $\bar f_{km}$ is trivially singular if $\Psi$ has no dependence on $\alpha$. In order to solve the obstruction problem to the construction of the Lagrangian for the rotational system, the matrix $\bar f$ must be nonsingular. Then, $\Psi$ necessarily depends on $\alpha$ variable. Thus, a vector $\bar\nu = \pmatrix{\nu_i & b & d}$, with $\nu_i\equiv\nu_i(\alpha)$ and $b$ and $d$ are arbitrary constants, when contracted with the symplectic matrix does not  generate a null value, {\it i.e.}, $\bar\nu \cdot {\bar f} \not= 0$. This relation generates a set of differential equations that allows the determination of $\Psi$. 
At this point, it is oportune to comment the arbitrariness of the vector $\bar\nu$. This arbitrariness opens up the possibility to determine different nonsingular symplectic matrix, then, we can obtain a set of different Lagrangian descriptions related to them. 

Now, it is necessary to determine $G$ in order to obtain dynamical Lagrangian for a rotational system. To this end, we impose that the contraction of the zero-mode, $\bar\nu$, with the gradient of symplectic potential, $\bar H(\bar\xi)$, generates a null value, given by $\bar\nu^k \frac{\delta\bar H(\bar\xi)}{\delta\xi^k} = 0$. From this condition and using the expressions for $\bar H$ and $G$, a set of general differential equations is obtained as
\be
\label{0140}
0 = \nu^i \frac{\delta H(\xi)}{\delta\xi^i} + b \sum_{n=1}^\infty\frac{\delta{\cal G}^{(n)}(\bar\xi)}{\delta\alpha} + d\sum_{n=1}^\infty \frac{\delta{\cal G}^{(n)}(\bar\xi)}{\delta\beta}.
\ee
>From this general equation, all correction term ${\cal G}^{(n)}(\bar\xi)$ can be computed just solving a differential equation obtained after the collection of all terms in Eq.(\ref{0140}) that belong to the same order in $\eta$.

\section{Rotational fluid mechanics}

In this section, the obstruction problem to construct both a canonical formalism and the Lagrangian description for rotational fluid mechanics will be described. Let us consider a inviscid, isentropic and compressible fluid, whose dynamics is governed by the continuity and Euler equations, which are read as $\frac{\partial \rho(t,\vec r)}{\partial t} + \bigtriangledown \cdot (\rho(t,\vec r)\cdot {\vec v}(t,\vec r)) = 0$, and $\frac{\partial {\vec v}(t,\vec r)}{\partial t} + {\vec v}(t,\vec r)\cdot \bigtriangledown {\vec v}(t,\vec r) = {\vec f}(t,\vec r)$, respectively, where ${\rho}(t,\vec r)$ and ${\vec v}(t,\vec r)$ denote mass density  and velocity field, respectively. Here, $\rho(t,\vec r){\vec v}(t,\vec r)$ is the current and ${\vec f}(t,\vec r)$ is the force, which will kept arbitrary for the time being.
It is well known that a dynamical system is powerfully presented from a canonical formulation. Due to this, it is important to remark that the above equations can be obtained by Poisson-bracketing the fields $\rho(t,\vec r)$ and ${\vec v}(t,\vec r)$ with the following Hamiltonian ${\cal H} = \frac 12 \rho v^2 + V(\rho)$, with $V(\rho)$ being an interactive potential. As a consequence, the Hamilton's equations of motion are $\frac{\partial \rho(t,\vec r)}{\partial t} = \{H, \rho(t,\vec r)\}$ and $\frac{\partial {\vec v}(t,\vec r)}{\partial t} = \{H,{\vec v}(t,\vec r) \}$, providing that the nonvanishing Poisson brackets among the fields must be taken as $\{v^i(\vec r), \rho(\vec r^\prime)\} = \frac{\partial\delta(\vec r - \vec r^\prime)}{\partial x_ i}$ and $\{v^i(\vec r), v^j((\vec r)\} = - \frac{\omega_{ij}(\vec r,\vec r^{\prime})}{\rho(\vec r)}\delta(\vec r - \vec r^\prime)$, where the vorticity $\vec w$ is $\omega_{ij}(\vec r,\vec r^\prime) = \frac{\partial v^j(\vec r^\prime)}{\partial x^i} - \frac{\partial v^i(\vec r)}{\partial x^{\prime j}}$.

In the symplectic canonical formulation of the rotational fluid mechanics,  we note that $f^{ij}$ has no inverse and, then, the symplectic two-form $f_{ij}$ does not exist. Therefore, the existence of a such constant $C$ creates an obstruction in the inversion of the symplectic matrix and, as a consequence, a canonical Lagrangian formulation for rotational fluid mechanics is lacking. To overcome this kind of problem and then neutralize the obstruction, the Clebsch parameterization process is usually implemented. However, in next section, in order to neutralize the obstruction problem, we apply the Clebsch-symplectic formalism process to the rotational fluid mechanics.

\section{Canonical Lagrangian formulation of the rotational fluid}

We begin considering the irrotational fluid mechanics, whose dynamics is governed by the following first-order Lagrangian, ${\cal L} = - \rho\dot\theta - \frac 12 \rho(\partial_i\theta)(\partial^i\theta) - V(\rho)$, where $\rho$ is the mass density and $\theta$ is the velocity potential. Here, the vorticity vanishes. Now, the Clebsch-symplectic parameterization process starts. Following the prescription of the Clebsch-symplectic formalism process, the Lagrangian is rewritten as ${\cal L} = - \rho\dot\theta + \Psi\dot\beta - \frac 12 \rho(\partial_i\theta)(\partial^i\theta) - V(\rho) - G$. Recalling that the arbitrary functions, $\Psi$ and $G$, will be determined later and that these functions present dependence on $\bar\xi^k=(\rho,\theta,\alpha,\beta)$. At this point, it is important to mention that the $\Psi$ function is determined in order to become the symplectic matrix nonsingular. Recalling that $\Psi$ has, necessarily, a dependence on $\alpha$, a vector to be chosen also must has a dependence on $\alpha$. To put our result in perspective with the usual Clebsch parameterization process\cite{CCL,jackiw}, we make a \lq\lq educated guess\rq\rq for $\nu$, namely, $\nu = \pmatrix{0 & \alpha & 0  & - 1}$.
Now, contracting this zero-mode with the symplectic matrix, and using the condition $\bar\nu \cdot {\bar f} \not= 0$, we get a set of differential equations for $\Psi$, and we obtain  $\Psi = - \alpha\rho$. The next step is the calculation of $G$. Using the Eq.(\ref{0140}), we obtain ${\cal G}^{(1)} = \alpha\rho(\partial^i\theta)(\partial^i\beta)$, ${\cal G}^{(2)} = \frac 12\alpha^2\rho(\partial^i\beta)(\partial^i\beta)$. As the second correction term has no dependence on $\theta$, all correction terms ${\cal G}^{(n)}$ with $n\geq 3$ are null. Hence, the Clebsch-symplectic parameterization process ended and  the canonical Lagrangian for the rotational fluid is obtained as being
\ba
\label{0330}
{\cal L} &=& - \rho(\dot\theta + \alpha\dot\beta) - \frac 12 \rho(\partial_i\theta + \alpha\partial_i\beta)^2 - V(\rho).
\ea
This reproduces the usual result obtained using Clebsch parameterization\cite{CCL,jackiw}.
Further, it is important to note for a remarkable and new result is demonstrated in this paper: it is possible to obtain different canonical Lagrangian descriptions for the rotational system, dynamically equivalent to the one given in Eq.(\ref{0330}), just defining a different vector $\nu$.

\section{Conclusion}

This work is devoted to solve the obstruction problem to the construction of the canonical Lagrangian formulation for rotational systems. This is done in the framework of the symplectic formalism and with the introduction of extra variables. We have demonstrate that there is not an unique way to neutralize the obstruction to construction of a canonical Lagrangian formulation for rotational systems, instead, there are different ways to neutralize the obstruction and, subesequently, different canonical Lagrangian descriptions, but equivalent, for rotational systems. Further, we have reproduced the usual dynamical Lagrangian for the rotational fluid mechanics, acquired by using the Clebsch parameterization, and suggested that there are a set of Lagrangian descriptions equivalent to the usual.

\noindent ACKNOWLEDGMENTS: This work is partially supported by CNPq and FAPEMIG, Brazilian Research Agencies.


\begin{thebibliography}{99}
\bibitem{LL}L. Landau and E. Lifshitz, {\it fluid Mechanics} (2nd ed. Pergamon, Oxford UK 1987).
\bibitem{HB}M. Bordemann and J. Hoppe, Phys. Lett. B317, 315 (1993); A. Jevicki, Phts. Rev. D57, 5955 (1988).
\bibitem{JB}D. Bazeia and R. Jackiw, Ann. Phys. 270, 246 (1988); D. Bazeia, Phys. Rev. D59, 085007 (1999). 
\bibitem{jackiw}R. Jackiw, {\it A Particle Field Theorist's Lectures on Supersymetry, Nonabelian Fluid Mechanics and D-branes.},e-Print Archive: physics/0010042. 
\bibitem{CCL}C.C. Lin, {\it International School of Physics E. Fermi (XXI)}, G. Careri, ed. (Academic Press, New York, NY 1963).
\bibitem{BW}J.Barcelos-Neto and C.Wotzasek, Mod.Phys.Lett.A 7, 1172 (1992); \\Int.J.Mod.Phys.A7 4981 (1992).
 
\end{thebibliography}
\end{document}